\documentclass[
  manuscript=article,  
  layout=publish,  
  year=20XX,
  volume=XX,
]{style}

\StartPage{1}

\received {22 June 2025}
\accepted {16 September 2025}
\published{XX XXX 2025}

\usepackage[english]{babel} 
\usepackage{blindtext}
\usepackage[right]{lineno}
\usepackage{caption}
\usepackage{subcaption}
\usepackage{isotope}

\title{Close Binary Open Cluster Systems: A Gaia DR3 Perspective on Alessi 36 and Collinder 135}

\author{H. Karagöz}
\affiliation{Istanbul University, Institute of Graduate Studies in Science, Programme of Astronomy and Space Sciences, 34116, Istanbul, Turkey}

\author{D. C. \c{C}ınar}
\affiliation{Istanbul University, Institute of Graduate Studies in Science, Programme of Astronomy and Space Sciences, 34116, Istanbul, Turkey}

\author{S. Ta\c{s}demir}
\affiliation{Istanbul University, Institute of Graduate Studies in Science, Programme of Astronomy and Space Sciences, 34116, Istanbul, Turkey}

\author{R. Canbay*}
\affiliation{Istanbul University, Faculty of Science, Department of Astronomy and Space Sciences, 34119, Beyaz\i t, Istanbul, Turkey \\ Email: rmzycnby@gmail.com}


\begin{document}

\begin{abstract}
  In this study, we examined the nearby open clusters (OCs) Alessi~36 and Collinder~135. Their spatial proximity was assessed using photometric and astrometric data from \textit{Gaia} DR3. Likely member stars were identified based on a membership probability threshold ($P \geq 0.5$), yielding 230 members for Alessi~36 and 342 members for Collinder~135. The mean proper-motion components ($\mu_{\alpha}\cos\delta, \mu_{\delta}$) were determined as $(-9.681 \pm 0.072,\ 7.021 \pm 0.081)$~mas~yr$^{-1}$ for Alessi~36 and $(-10.061 \pm 0.083,\ 6.256 \pm 0.095)$~mas~yr$^{-1}$ for Collinder~135. The parallax-based distances ($d_{\varpi}$) were found to be $278 \pm 7$ and $299 \pm 11$~pc, while their estimated ages ($t$) were $39 \pm 5$ and $36 \pm 5$~Myr for Alessi~36 and Collinder~135, respectively. The Galactic orbital analysis of Alessi~36 and Collinder~135 reveals nearly circular orbits with low eccentricities and minor fluctuations in their apogalactic and perigalactic distances. The maximum heights they attain above the Galactic plane are $0.035 \pm 0.001$~kpc for Alessi~36 and $0.074 \pm 0.003$~kpc for Collinder~135, respectively, supporting their classification as part of the young stellar disc population.
\end{abstract}

\section{Introduction}

Star formation occurs over a wide range of spatial and temporal scales, giving rise to both gravitationally bound clusters and more diffuse stellar associations. These structures often exhibit hierarchical organization, reflecting the underlying physical processes that govern star formation and early cluster evolution {[}1,5{]}. Among these stellar systems, open clusters (OCs) arise from the fragmentation and gravitational collapse of dense regions within giant molecular clouds (GMCs). While some OCs may form in relative isolation, accumulating stars from their neighborhood, others appear to form as part of larger stellar complexes, commonly referred to as primordial groups. Observational studies provide compelling evidence that OCs forming in close proximity often share common kinematic properties, suggesting a common origin. Analyzing spatial and kinematic relationships among young OCs enhances our understanding of hierarchical star formation processes in the Milky Way and improves our knowledge of their long-term dynamical
evolution.

The launch of ESA's \emph{Gaia} mission has revolutionized the field of precision astrometry, allowing the study of Galactic clusters with unprecedented accuracy, and facilitating the discovery of previously unknown OCs {[}6, 7{]}. The release of \emph{Gaia} Data Release 3 (\emph{Gaia} DR3) on 13 June 2022 marks a major milestone in this effort, providing positional and photometric measurements for about 1.8 billion sources, along with trigonometric parallax ($\varpi$), proper motion
components (($\mu_{\alpha}\cos\delta, \mu_{\delta}$), and
\emph{BP}-\emph{RP} colours for nearly 1.5 billion sources {[}6{]}.

The substantial improvements in astrometric precision, especially in PMs, compared to \emph{Gaia} DR2 have opened new avenues for revisiting the OC population and refining its fundamental properties. The new findings show that some open clusters are close to each other and may have gravitational interactions {[}15,19{]}. These systems, known as close binary open clusters (CBOCs), provide unique opportunities to study the interaction between clusters and their surrounding Galactic environment. The study of CBOCs provides key insights into cluster dynamics, their interactions, and the role of the Galactic gravitational field in establishing their long-term stability.

\begin{table}[ht]
\centering
\footnotesize
\caption{Astrophysical parameters of Alessi~36 and Collinder~135 from the literature.}
\label{tab:alessi_collinder}
\resizebox{\textwidth}{!}{%
\begin{tabular}{cccccccc}
\hline
$E(B-V)$ & $d$ & [Fe/H] & $t$ & $\mu_{\alpha}\cos\delta$ & $\mu_{\delta}$ & $v_{r}$ & References \\
(mag) & (pc) & (dex) & (Myr) & (mas yr$^{-1}$) & (mas yr$^{-1}$) & (km s$^{-1}$) & \\
\hline
\multicolumn{8}{c}{\textbf{Alessi 36}} \\
\hline
0.032 & 276 & --- & 32 & $-9.703 \pm 0.199$ & $6.982 \pm 0.216$ & --- & [07] \\
$0.045 \pm 0.020$ & $275 \pm 1$ & --- & $16 \pm 9$ & $-9.669 \pm 0.016$ & $7.046 \pm 0.089$ & $14.54 \pm 1.82$ & [08] \\
--- & --- & --- & 40 & --- & --- & --- & [10] \\
--- & $273 \pm 2$ & --- & --- & 14.2 & -7.4 & 11.6 & [11] \\
--- & --- & --- & --- & $-10.25 \pm 0.05$ & $-5.98 \pm 0.05$ & $16.7 \pm 1.5$ & [12] \\
\hline
\multicolumn{8}{c}{\textbf{Collinder 135}} \\
\hline
0.003 & 292 & --- & 26 & $-9.975 \pm 0.445$ & $6.157 \pm 0.350$ & --- & [07] \\
$0.034 \pm 0.01$ & $295 \pm 1$ & --- & $30 \pm 23$ & $-10.077 \pm 0.019$ & $6.23 \pm 0.017$ & $13.16 \pm 1.14$ & [08] \\
--- & --- & --- & 40 & --- & --- & --- & [10] \\
--- & $294 \pm 2$ & --- & --- & 13.6 & -6.9 & 6.6 & [11] \\
--- & --- & --- & --- & $-9.92 \pm 0.05$ & $-6.47 \pm 0.06$ & $17.4 \pm 1.3$ & [12] \\
$0.058 \pm 0.02$ & $307 \pm 8$ & $0.111 \pm 0.058$ & $50 \pm 6$ & $-9.977 \pm 0.491$ & $6.211 \pm 0.453$ & $16.513 \pm 2.254$ & [13] \\
$0.07 \pm 0.01$ & $310 \pm 8$ & $0.10 \pm 0.002$ & $38 \pm 2$ & --- & --- & $2.87 \pm 0.17$ & [12] \\
--- & --- & --- & --- & $-9.280 \pm 0.027$ & $5.57 \pm 0.22$ & --- & [13] \\
-0.06 & 630 & --- & --- & --- & --- & --- & [14] \\
\hline
\end{tabular}}
\end{table}

Based on on the analysis of {[}20{]}, this study provides an in-depth investigation of the binary open cluster Collinder 135 and Alessi 36 through a comprehensive astrometric, astrophysical, and dynamic orbit analysis. While previous work has suggested possible gravitational interaction between these clusters, our study improves their characterization by using high-precision astrometric measurements and advanced orbital modelling techniques. By thoroughly examining their kinematics, structural properties, and orbits, we aim to determine whether they form a physically bound binary open cluster system or merely exhibit a random alignment. This refined analysis will advance our understanding of the dynamical evolution of pair open clusters in the Galactic field by elucidating the physical mechanisms that govern their formation, mutual interactions, and long-term stability.

Collinder 135 and Alessi 36 have been the subject of various studies aimed at determining their key astrophysical properties, including color excess, proper motion components, ages, metallicities, and distances. These clusters provide an ideal environment for studying their dynamics, with potential interactions influencing their evolution. A summary of these studies and their results is given in Table 1.

The remaining sections of this paper are structured as follows: Section 2 describes the data sources and selection criteria used in this study. Section 3 outlines the methodology used to derive the fundamental astrometric, astrophysical, and kinematical parameters of the clusters and presents the results. Section 4 provides a comprehensive summary of the study, consolidating the main conclusions drawn from our analysis.

\section{Data}

The photometric and astrometric investigation of Collinder~135 and Alessi~36 was conducted using data obtained from the \emph{Gaia} DR3 catalogue {[}8{]}. \emph{Gaia} DR3 provides high-precision astrometric, photometric, and radial velocity measurements for over 1.8 billion sources, including trigonometric parallax ($\varpi$), proper-motion components ($\mu_{\alpha}\cos\delta$, $\mu_{\delta}$), and integrated $G$, $G_{\rm BP}$, and $G_{\rm RP}$ magnitudes for approximately 1.5 billion stars. Additionally, \emph{Gaia} DR3 includes radial velocities for $\sim$33 million bright stars ($G \leq 14$~mag) and astrophysical parameters such as effective temperature, extinction, and metallicity for millions of sources. 

These two OCs are located in a relatively sparse region of the Milky Way, providing valuable opportunities for detailed observational studies {[}21{]}. Collinder~135 is located at $\alpha = 07^{\rm h}17^{\rm m}26^s9$ and $\delta = -37^{\circ}02'38''$ (J2000.0), with Galactic coordinates of $l = 248.9884$ and $b = -11.2010$, while Alessi~36 is located at $\alpha = 07^{\rm h}07^{\rm m}08^s6$ and $\delta = -37^{\circ}43'44''$ (J2000.0), corresponding to Galactic coordinates of $l = 248.7423$ and $b = -13.3458$. 

To ensure a comprehensive examination, all stars within a $90'$ radius from the centre of each cluster were included in the study. As a result, a total of 300\,707 stars and 215\,885 stars in the magnitude range $6 < G \leq 23$~mag were identified for Collinder~135 and Alessi~36, respectively. These datasets form the basis for the subsequent membership analysis and derivation of the clusters' fundamental parameters.

\section{Method and Results}
\subsection{Cluster Membership Determination: UPMASK} 

The UPMASK (Unsupervised Photometric Membership Assignment in Stellar Clusters) program {[}22{]} was used to calculate the membership probabilities of stars in the Alessi 36 and Collinder 135 regions. This program utilizes the k-means clustering algorithm, which groups parameters with similar characteristics within a dataset and assigns the most appropriate probability values. The k-means values are user-defined and can vary between 6 and 30, depending on the cluster morphology, to provide the optimal solution. The algorithm works by dividing the data into a set number of clusters (k), assigning each point to the nearest cluster center, and updating the centers iteratively until the best grouping is achieved.

Following the UPMASK {[}22{]} guidelines, we explored $k \in (6,30)$ in the five-dimensional space used for membership (position and astrometry). For each $k$ we computed three diagnostics: (i) the mean silhouette coefficient in the ($\mu_{\alpha}\cos\delta$, $\mu_{\delta}$, $\varpi$) subspace as an internal validity metric; (ii) a surface-density contrast between members and surrounding field within the cluster half-number radius as a compactness proxy; and (iii) a stability score defined as the Jaccard index between member lists obtained at adjacent $k$ values. The solutions become stable for $k \gtrsim 20$; derived bulk parameters (mean proper motions, mean parallax, distances) vary within their $1\sigma$ uncertainties across this regime, while the intersecting fraction of members between neighboring $k$ values exceeds 0.90 for both clusters. We therefore adopt $k=27$, which provides near-maximal silhouette and stability without over-fragmenting the data.

A membership probability threshold of $P \geq 0.5$ was adopted. This value is commonly used in UPMASK-based studies and provides a reliable separation of cluster members from field stars {[}7,8{]}. To test sensitivity, we also examined thresholds of $P \geq 0.3$ and $P \geq 0.7$. The former led to noticeably higher field-star contamination, while the latter removed genuine members even in the cluster center. The adopted $P \geq 0.5$ criterion therefore represents the most balanced and literature-consistent choice. Consequently, the number of highly probable members was found to be 230 for Alessi~36 and 342 for Collinder~135.

\subsection{Astrometric Parameters} 

To analyze the distribution of stars with high membership probability ($P \geq 0.5$) in the proper motion space relative to field stars and to determine the mean proper motion components, vector point diagrams (VPDs) were constructed for the Alessi~36 and Collinder~135 clusters (Figure~1; right panel). Additionally, the sky-projected vectorial motions of these stars in equatorial coordinates were examined. The results of this analysis are presented in Figure~1 (left panel), which shows the VPDs and the equatorial coordinate distributions for both clusters, clearly distinguishing high-probability members and field stars. In this figure, stars with membership probabilities $P \geq 0.5$ and those with $0 < P < 0.5$ are represented by colored and grey circular symbols, respectively. The analysis indicates that Alessi~36 and Collinder~135 are significantly influenced by field stars.

\begin{figure}[h]
\centering
\includegraphics[width=0.75\linewidth]{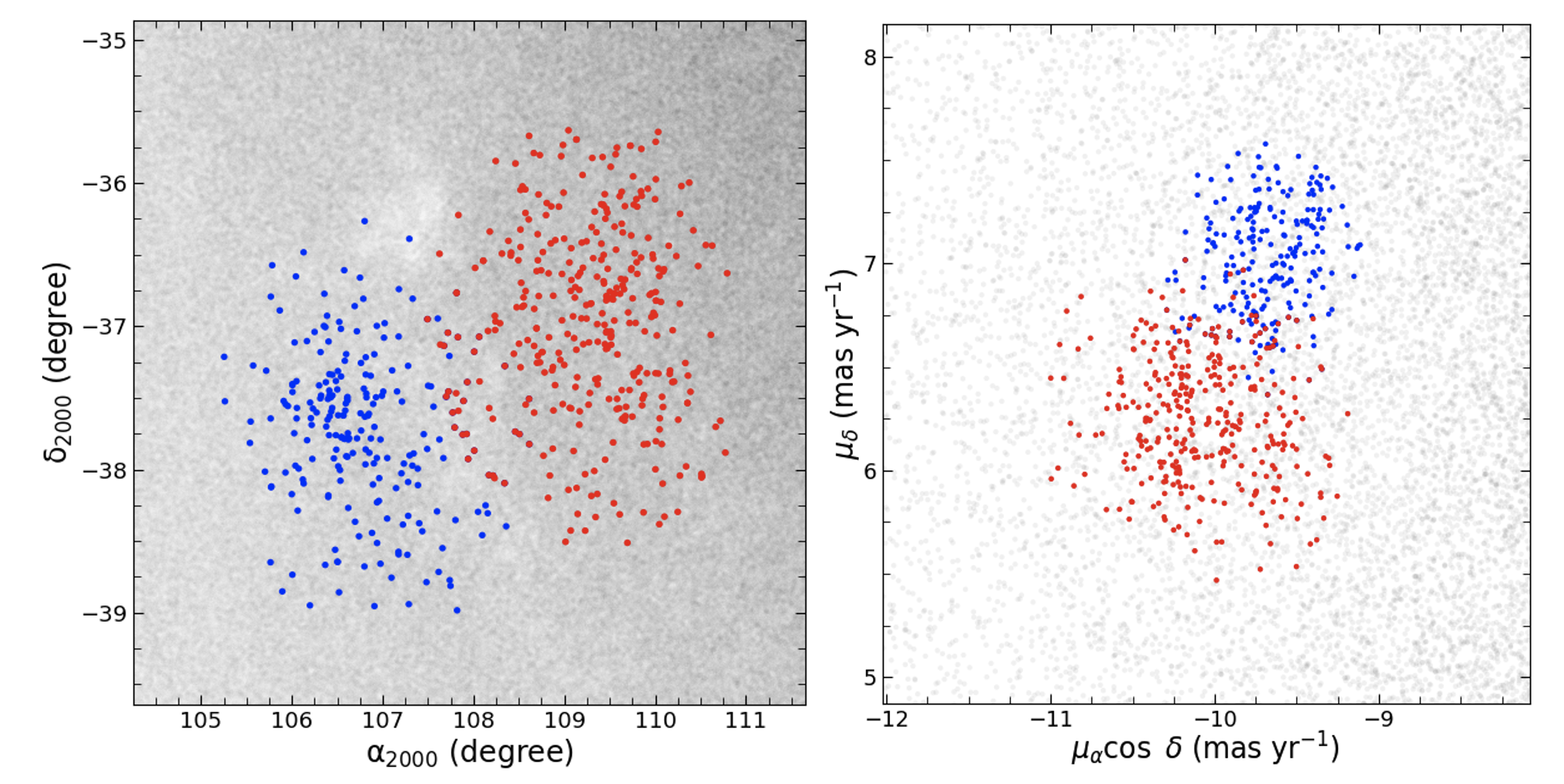}
\caption{The image of Alessi 36 (blue dots) and Collinder 135 (red dots) distributed on the equatorial coordinate plane (left panel). VPD of the most likely member stars belong to the open clusters (right panel). The background grey points represent field stars.}
\end{figure}

A detailed examination of the high-probability cluster members ($P \geq 0.5$) reveals that Alessi~36 and Collinder~135 show a significantly higher stellar density near their centers compared to their outskirts, indicating a dense central stellar distribution, and they appear as a binary system in the sky. The mean proper motion components ($\langle \mu_{\alpha}\cos\delta, \mu_{\delta} \rangle$) for these clusters were determined by calculating the median values of parameters with $P \geq 0.5$. The resulting values for Alessi~36 and Collinder~135 are $(-9.703 \pm 0.004, 6.982 \pm 0.004)$ and $(-9.975 \pm 0.006, 6.157 \pm 0.007)$~mas~yr$^{-1}$, respectively (see Table~2).

Furthermore, the mean trigonometric parallaxes ($\langle \varpi \rangle$) of both clusters were computed. Histograms of the trigonometric parallax measurements for stars with membership probability $P \geq 0.5$ were generated, and a Gaussian function was fitted to the distribution. To ensure data reliability, we imposed a relative parallax error constraint of $\sigma_{\varpi}/\varpi \leq 0.2$. The mean parallaxes of Alessi~36 and Collinder~135 were thus determined as 3.601 and 3.348~mas, respectively. These values were then converted to heliocentric distances using the relation $d$~(pc) = 1000 / $\varpi$~(mas), yielding distances of $278 \pm 7$ and $299 \pm 11$~pc for Alessi~36 and Collinder~135, respectively.

We note that direct inversion of parallaxes can introduce biases (e.g., Lutz--Kelker effect) for distant stars with large relative uncertainties. However, for the nearby clusters studied here (around 300~pc) and with the applied condition $\sigma_{\varpi}/\varpi \leq 0.2$, such biases are negligible and the results are robust {[}41,43{]}. Moreover, the obtained distances are consistent with those derived from isochrone fitting (Section~3.3), further supporting the reliability of this approach. While Bayesian distance estimates {[}42{]} could serve as an alternative, we adopted the direct method here for its simplicity and consistency under our selection criteria.

As illustrated in Figure~2, the trigonometric parallax histograms of Alessi~36 and Collinder~135 are presented together with the Gaussian function (black dashed line), which has been fitted to the distributions. The astrometric parameters derived for both clusters are summarized in Table~2.

\begin{figure}[h]
\centering
\includegraphics[width=0.4\linewidth]{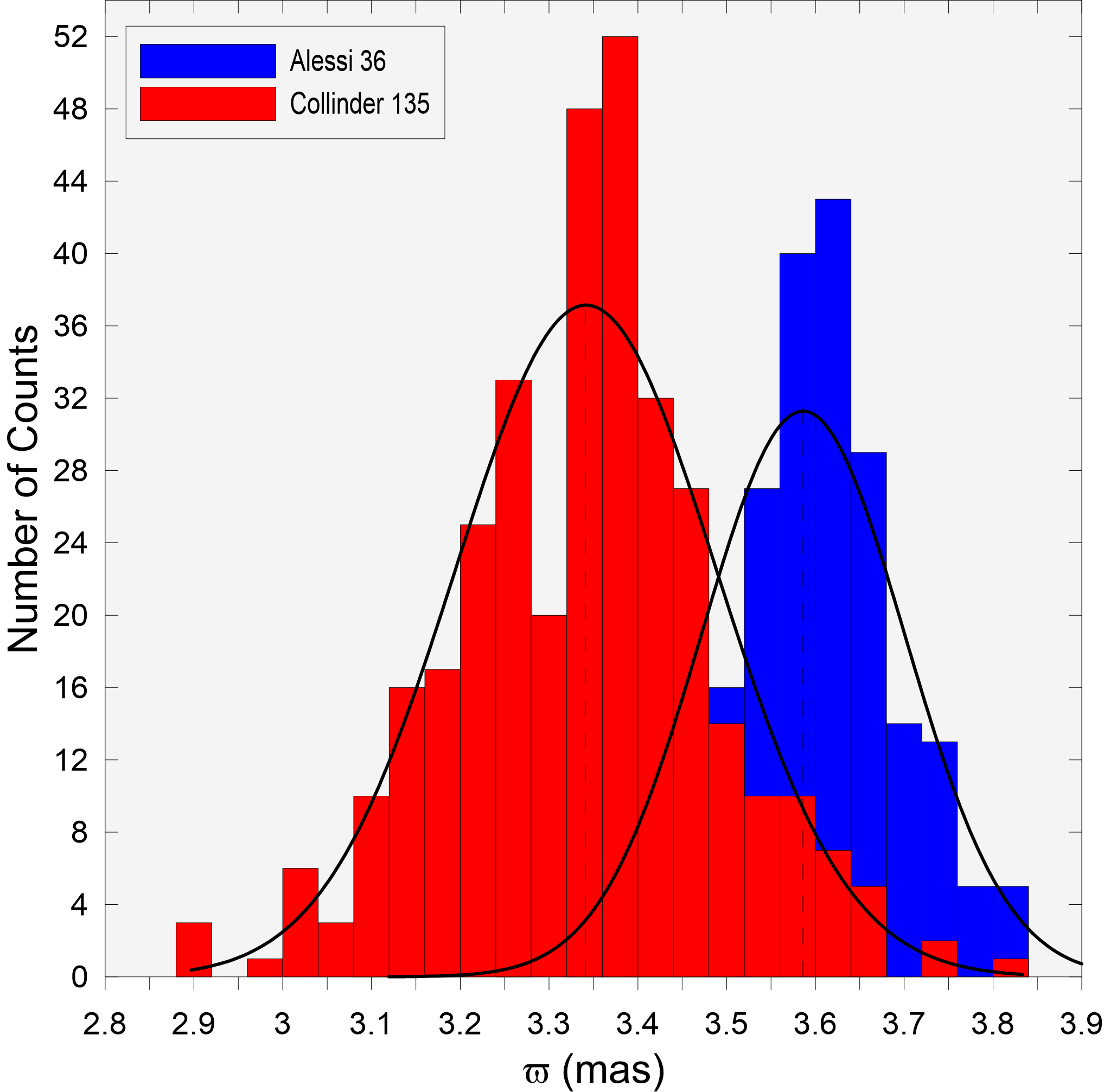}
\caption{The figure displays the trigonometric parallax
histograms based on \textit{Gaia} DR3 data, with black solid lines indicating the median of the Gaussian fits to the parallax distributions of Alessi 36 (blue bars) and Collinder 135 (red bars).}
\end{figure}

\subsection{Fundamental Astrophysical Parameters}

The interstellar reddening of open clusters is of critical importance in providing insights into the extent of dust extinction affecting their observed photometric properties {[}18,26,28{]}. Utilizing data from the \emph{Gaia} mission, the reddening values for Alessi~36 and Collinder~135 were determined using the color excess in the \emph{Gaia} photometric bands, specifically $E(G_{\rm BP} - G_{\rm RP})$. For Alessi~36, the derived reddening value is $E(G_{\rm BP} - G_{\rm RP}) = 0.104 \pm 0.027$~mag, indicating a moderate level of extinction within its Galactic environment. In contrast, Collinder~135 exhibits a lower reddening value of $E(G_{\rm BP} - G_{\rm RP}) = 0.065 \pm 0.022$~mag, consistent with its location in a region with relatively low interstellar dust content. These findings suggest that although both clusters experience some degree of interstellar extinction, the effect is more pronounced in Alessi~36 compared to Collinder~135. Despite their close proximity on the sky, the differing reddening values likely reflect variations in dust distribution or differences in their distances.

The reddening values of Alessi~36 and Collinder~135 were also derived by fitting the PARSEC isochrones to the observed \textit{Gaia} DR3 color--magnitude diagrams (CMDs), with the color excess determined in the Gaia photometric system as $E(G_{\rm BP} - G_{\rm RP})$ and shown in Figure~3. For Alessi~36, the derived value is $E(G_{\rm BP} - G_{\rm RP}) = 0.104 \pm 0.027$~mag, while for Collinder~135 it is $E(G_{\rm BP} - G_{\rm RP}) = 0.065 \pm 0.022$~mag. Converting to the \emph{UBV} system using the relation $E(G_{\rm BP} - G_{\rm RP}) = 1.41 \times E(B-V)$, we obtained $E(B-V) = 0.074 \pm 0.019$~mag and $E(B-V) = 0.046 \pm 0.016$~mag, respectively. These results are consistent with earlier estimates reported in the literature {[}23,27,29,30{]}, which place the reddening of both clusters in the range $E(B-V) \approx 0.03$--$0.07$~mag. The agreement confirms that our derived reddening values are robust and reliable for subsequent distance and age determinations.

\begin{figure}[h]
\centering
\includegraphics[width=0.4\linewidth]{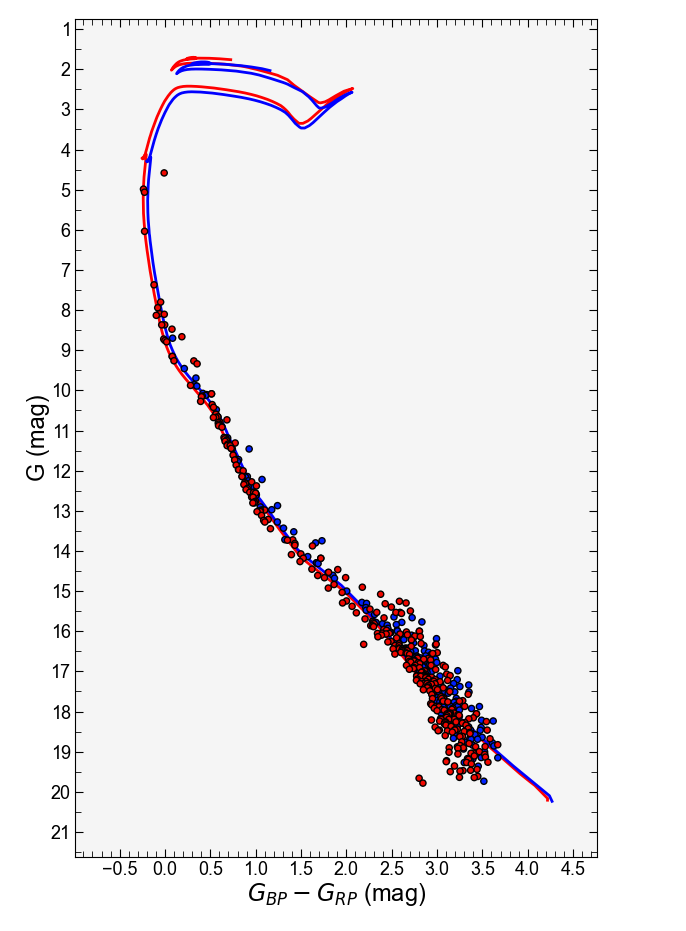}
\caption{Color-magnitude diagram of the OCs of clusters Collinder 135 (red) and Alessi 36 (blue).}
\end{figure}

The determination of metallicity in open clusters is important for understanding their chemical composition, formation environment, and evolutionary status. In instances where spectroscopic data is not available, photometric methods emerge as a reliable approach for estimating the metal abundance of stellar populations by using color indices and empirical calibrations. Photometric metallicity values for Alessi~36 and Collinder~135 have been derived based on \emph{Gaia} observations and calibrated photometric relations. To select isochrones for determining age and distance modulus, [Fe/H] of each open cluster was converted into the mass fraction $z$. This conversion used formulas adapted from Bovy\footnote{https://github.com/jobovy/isodist/blob/master/isodist/Isochrone.py} based on the PARSEC {[}45{]} models. The process involves calculating an intermediate value $z_x$ and then obtaining $z$, with the solar metallicity fixed at $z_{\odot}=0.0152$:
\[
\begin{matrix}
z_x = 10^{[{\rm Fe/H}] + \log \left( \frac{z_{\odot}}{1 - 0.248 - 2.78 \times z_{\odot}} \right)} \\
z = \frac{z_x (1 - 0.2485)}{1 + 2.78 \times z_x} \\
\end{matrix}
\]

Applying this to Alessi~36 and Collinder~135 yielded mass fractions of 0.0152 and 0.01195, respectively. The results indicate that both clusters exhibit metallicities close to the solar value, consistent with their locations in the Galactic disk. The estimated metallicity values are [Fe/H] = $0.00 \pm 0.05$ and $-0.01 \pm 0.04$~dex, respectively. Nonetheless, our result for Collinder~135 is consistent with the spectroscopic value reported by {[}12{]}, and both clusters fall within the typical metallicity range of young open clusters in the solar neighbourhood. The conclusion that Alessi~36 has a metallicity close to solar is thus supported by the photometric calibrations and indirectly by the agreement with spectroscopic findings for Collinder~135.

A comparative analysis with other open clusters of similar age and location suggests that Alessi~36 and Collinder~135 share chemical properties typical of intermediate-age clusters in the solar neighbourhood. Determining the distance and age of open clusters is essential for understanding their evolutionary status, formation environment, and role in the Galactic structure. Utilizing \emph{Gaia}'s high-precision astrometric data, the distances of Alessi~36 and Collinder~135 were estimated as $279 \pm 7$ and $296 \pm 6$~pc, respectively. These values indicate that both clusters are relatively nearby and situated within the Galactic disk, making them valuable objects for studying stellar evolution and kinematics in the solar neighborhood. The distances quoted here are the isochrone-based values ($d_{\rm iso}$), adopted as our fiducial distances. They are fully consistent within $1\sigma$ with the parallax-based distances ($d_{\varpi}$) reported in Section~3.2. The small present-day $|Z|$ (-64 and -57~pc) and low $Z_{\rm max}$ (35 and 74~pc) further indicate membership in the young thin disc, in line with the concentration of young open clusters toward the Galactic plane reported in previous \emph{Gaia} studies {[}7,21{]}.

The ages of the clusters were estimated through isochrone fitting of \emph{Gaia} DR3 photometry using PARSEC models, yielding $39 \pm 5$~Myr for Alessi~36 and $36 \pm 5$~Myr for Collinder~135. These values indicate that both systems are relatively young, consistent with literature estimates for nearby open clusters of a similar type. At such ages, which are shorter than typical two-body relaxation times for clusters of comparable size, significant mass segregation or tidal disruption is not expected to have occurred yet. This supports the interpretation that both clusters are still in an early stage of their dynamical evolution. Their nearly coeval ages further suggest that they may have formed under comparable Galactic conditions or within similar star-forming environments, in line with previous studies of young disc clusters.

\subsection{Kinematic and Galactic Orbit Analyses}

The kinematical properties of open clusters are of crucial significance in providing insights into their dynamical evolution, Galactic motion, and possible origin within the Milky Way {[}24{]}. Utilizing data from the \emph{Gaia} astrometric and radial velocity surveys, the kinematic parameters of Alessi~36 and Collinder~135 were obtained. Orbital integrations were then carried out using the Galpy package {[}40{]}, adopting the MilkyWayPotential model consisting of a Miyamoto--Nagai disk, a Hernquist bulge, and an NFW halo. This axisymmetric potential is widely applied in open cluster studies and provides a realistic description of the solar neighbourhood dynamics. The orbits were integrated backward for 3~Gyr with a timestep of 1.5~Myr using a fourth-order Runge--Kutta scheme. Initial conditions were set using \emph{Gaia} DR3 astrometry (positions, parallaxes, proper motions) and \emph{Gaia} radial velocities, ensuring consistency with previous Gaia-based orbital analyses of young open clusters.

This encompassed their Galactic positions ($X$, $Y$, $Z$), radial velocity, Galactocentric distance ($R_{\rm gc}$), and velocity components relative to the Local Standard of Rest ($U$, $V$, $W$)$_{\rm LSR}$. Systemic radial velocities were computed as inverse-variance weighted means of $Gaia$ RVS measurements for probable members ($P \geq 0.5$) with iterative $3\sigma$ clipping. For Alessi~36, 55 stars yielded $v_r = 15.58 \pm 0.42$~km~s$^{-1}$, while 12 stars gave $v_r = 15.62 \pm 1.11$~km~s$^{-1}$ for Collinder~135. Heliocentric Galactic Cartesian coordinates were derived from $(l, b, d)$ using the standard transformations, yielding $(X, Y, Z) \approx (-98, -253, -64)$~pc for Alessi~36 and $(-104, -271, -57)$~pc for Collinder~135 (see Table~2). Galactocentric radii were calculated assuming $R_0 = 8.0$~kpc, giving $R_{\rm gc} = 8.10$ and 8.11~kpc, respectively, showing that both clusters lie at essentially the same Galactocentric distance. A right-handed coordinate system was adopted, with $U$ directed toward the Galactic center, $V$ along the direction of Galactic rotation, and $W$ toward the North Galactic Pole. Velocities were corrected to the Local Standard of Rest using $(U, V, W)_{\odot} = (8.5, 13.38, 6.49)$~km~s$^{-1}$ {[}44{]}. The results, summarized in Table~2, show $(U, V, W)_{\rm LSR} = (-12.39 \pm 0.63, +14.05 \pm 0.34, +0.06 \pm 0.21)$~km~s$^{-1}$ for Alessi~36 and $(-9.13 \pm 0.76, +6.53 \pm 0.95, -5.22 \pm 0.31)$~km~s$^{-1}$ for Collinder~135.

The current positions of the clusters are marked with filled circles in Figure~4, whereas their birth positions are indicated by filled triangles. The motion directions of Collinder~135 and Alessi~36 are represented by red and blue arrows, respectively. Furthermore, the clusters' orbits, including uncertainties in the input parameters, are depicted in orange dotted lines. The triangles correspond to the lower and upper error margins for the clusters' birth locations, respectively (middle panels). The similarity of these values suggests that the clusters may have formed under similar initial conditions or may have originated from the same star-forming region. The Galactic positions ($X$, $Y$, $Z$) and Galactocentric distances ($R_{\rm gc}$) further confirm this hypothesis, demonstrating that both clusters are situated at comparable distances from the Galactic center.

\begin{figure}[h]
\centering
\includegraphics[width=0.9\linewidth]{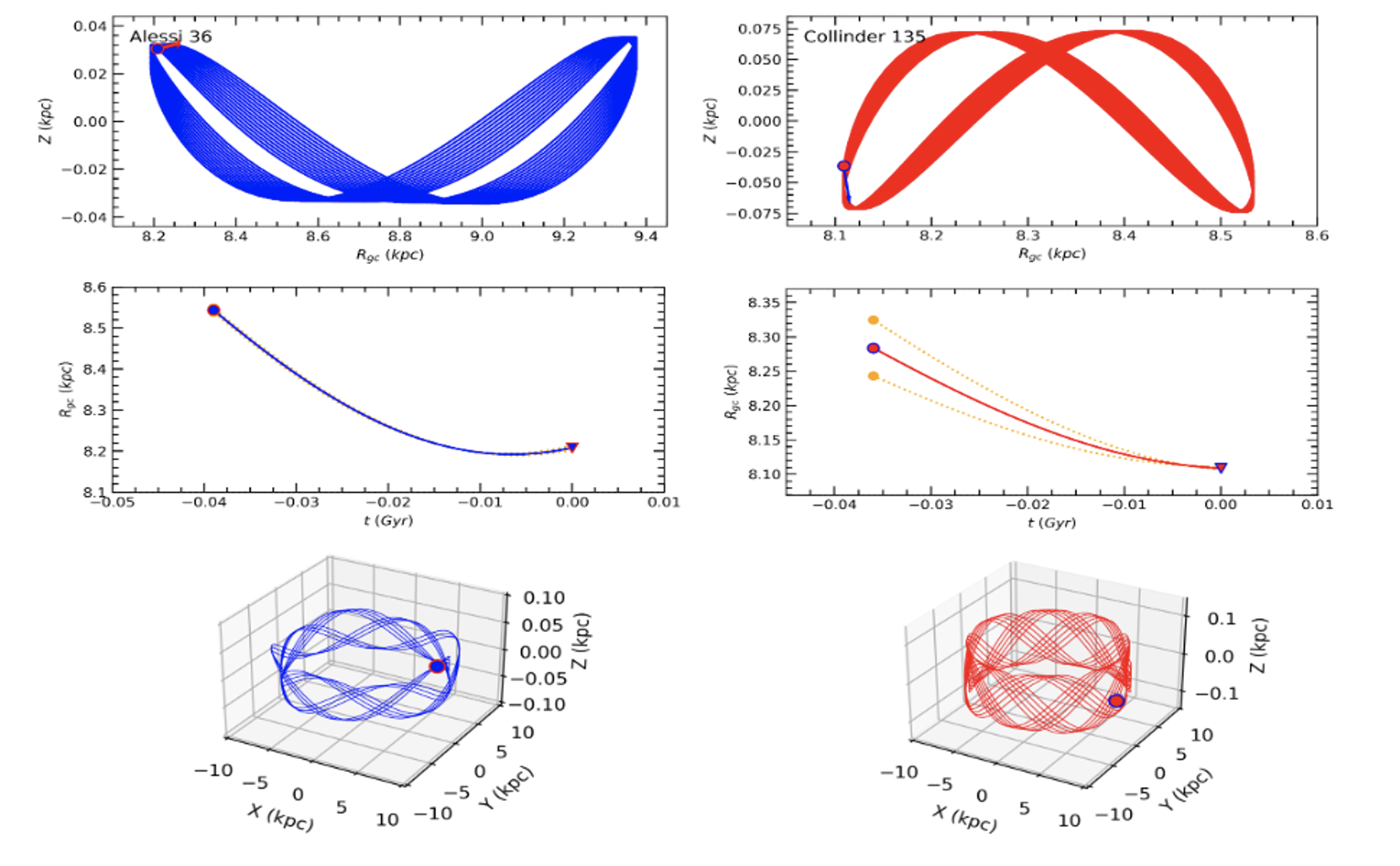}
\caption{The Galactic trajectories and birth radii of the Collinder 135 (red) and Alessi 36 (blue) clusters are illustrated across two separate panels. Upper panels show the $Z \times R_{\rm gc}$ plane, the middle panels present the $R_{\rm gc} \times t$, the circles and triangles represent birthplace and current locations, respectively. The bottom panels visualize the $X \times Y \times Z$ plane.}
\end{figure}

Figure~4 illustrates the Galactic trajectories and birth radii of Collinder~135 (red) and Alessi~36 (blue) clusters across two separate panels. Upper panels show the $Z \times R_{\rm gc}$ plane, the middle panels present $R_{\rm gc} \times t$, with circles and triangles representing current and birthplace locations, respectively. The bottom panels visualize the $X \times Y \times Z$ plane.

The velocity components relative to the Local Standard of Rest ($U$, $V$, $W$)$_{\rm LSR}$ provide additional insight into the orbital motion of the clusters {[}27,31,32{]}. These parameters are essential for understanding the clusters' trajectories within the Milky Way and their interactions with the Galactic potential. These modest velocities, together with the low orbital eccentricities ($e < 0.07$) and small vertical excursions ($Z_{\rm max} < 0.08$~kpc), are typical of young thin-disc open clusters {[}7,21{]} and confirm that both systems follow stable, nearly circular orbits within the Galactic disk, without evidence for recent strong gravitational perturbations or external interactions.

\section{Discussion}

In this study, we conducted a close binary cluster analysis of Alessi 36 and Collinder 135 via \emph{Gaia} DR3 data for photometric, astrometric, and kinematic evaluations. In the following subsections we discuss the results our analyses.

\subsection{Astrometric Parameters}

Based on membership probabilities $P \geq 0.5$, the most probable member stars of both clusters were identified using the machine learning k-means method. These stars formed the basis for deriving key astrophysical parameters and Galactic orbital characteristics. The age, distance, and reddening of the clusters were simultaneously determined using \emph{Gaia}-based CMDs. Using the vector point diagram (VPD) of most probable member stars, the mean proper-motion components $\mu_\alpha \cos \delta$, $\mu_\delta$ were estimated for Alessi 36 and Collinder 135 as $(-9.681 \pm 0.072, 7.021 \pm 0.081)$ and $(-10.06 \pm 0.083, 6.256 \pm 0.095)$~mas~yr$^{-1}$, respectively. The mean trigonometric parallaxes, $\langle \varpi \rangle$, obtained from Gaussian fits to parallax histograms constructed from the most likely stars are $3.601 \pm 0.086$ and $3.348 \pm 0.129$~mas for Alessi 36 and Collinder 135, respectively. Reference [8] presents the largest all-sky search for open clusters using \emph{Gaia} DR3, identifying 7,167 clusters with their astrometric, photometric, and kinematic parameters. As can be seen from Table 1, the proper motion components and trigonometric parallax values obtained from this study are in good agreement with the results presented by [8].

\subsection{Fundamental Astrophysical Parameters}

The main sequence fitting method is a fundamental technique for determining cluster parameters such as distance, age, and extinction by comparing observed CMDs with theoretical isochrones [7,12,17,23,26]. However, estimation of these parameters is often affected by degeneracies, since different combinations of parameters can produce similar fits, making precise determination challenging. In order to minimize this degeneracy, the trigonometric parallax distance and the metallicity value obtained from spectral observations in the literature were fixed in this stage. The color excess for both clusters was derived as $E(G_{\rm BP}-G_{\rm RP}) = 0.104 \pm 0.027$ and $E(G_{\rm BP}-G_{\rm RP}) = 0.065 \pm 0.022$~mag for Alessi~36 and Collinder~135, respectively. Using the relation $E(G_{\rm BP}-G_{\rm RP}) = 1.41 \times E(B-V)$ [33,34], the UBV-system color excesses were obtained as $E(B-V) = 0.074 \pm 0.019$~mag for Alessi~36 and $E(B-V) = 0.046 \pm 0.016$~mag for Collinder~135.

For Alessi~36, our derived distance ($279 \pm 7$~pc) and age ($39 \pm 5$~Myr) are in good agreement with the recent study by [36], which reported a distance of $281 \pm 9$~pc and an age of $41 \pm 6$~Myr based on $Gaia$ DR3 data. The proper motion components we obtained also overlap within uncertainties with their values, supporting the robustness of our membership determination. For Collinder~135, our results are broadly consistent with those of [37,38], who both reported distances near 300~pc and ages around 35--40~Myr. More recent determinations by [39,36] also provide similar results, with ages of $\sim 36$--38~Myr and distances close to 295--300~pc. The minor discrepancies in membership numbers and mean proper motions likely reflect differences in the adopted membership probability thresholds and astrometric filtering criteria. Overall, these comparisons confirm that our analysis is consistent with previous literature while providing refined estimates of the clusters' physical and kinematical properties.

\subsection{Kinematic Analyses}

Radial velocity measurements have an important place in performing kinematic analyses. In this study, radial velocity values measured with \emph{Gaia}'s radial velocity spectrometer (RVS) were used in kinematic analyses. When compared with other studies in the literature, it is seen that these measurement results are compatible with each other. Reference [21] constructed a comprehensive radial velocity catalogue with \emph{Gaia} DR2 data, giving radial velocity values for Alessi 36 and Collinder 135 as 16.80 $\pm$ 0.67 and 16.87 $\pm$ 0.23 km s$^{-1}$, respectively. Our measurements, 15.58 $\pm$ 0.42 km s$^{-1}$ for Alessi 36 and 15.62 $\pm$ 1.11 km s$^{-1}$ for Collinder 135, are slightly lower but remain broadly consistent within uncertainties (see Table 2). 

Galactic orbit analyses were conducted for both clusters. Using precisely determined ages, their orbits were traced backward in time to identify their birthplaces within the Galaxy, revealing that they originated in different regions. Alessi 36 is located at 8544 pc, while Collinder 135 is at 8284 pc (Figure 3). A review of the literature shows that [20] classified these clusters as pairs in their binary cluster catalogue. However, our kinematic analysis confirms that although they are binary clusters, they did not form simultaneously. Galactic orbit analysis revealed that both clusters follow low eccentricity nearly circular orbits within the young thin-disc population of the Milky Way [33,34,35].

\begin{table}[h!]
\scriptsize
\centering
\caption{Fundamental parameters of Alessi 36 and Collinder 135 compared with [8].}
\label{tab:fundamental}
\resizebox{\textwidth}{!}{%
\begin{tabular}{lccc}
\hline
\textbf{Parameters} & \textbf{Alessi 36} & \textbf{Collinder 135} & \textbf{Hunt and Reffert (2024)} \\
\hline
$(\alpha, \delta)_{J2000}$ (Sexagesimal) & 07 07 08.65, -37 43 44.5 & 07 17 26.89, -37 02 38.5 & 07 06 15.64, -37 35 45.49 / 07 17 37.34, -36 55 15.26 \\
$(l, b)_{J2000}$ (Decimal) & 248.742, -13.345 & 248.989, -11.201 & 248.544, -13.451 / 248.890, -11.116 \\
Members (P $\geq$ 0.5) & 230 & 342 & 198 / 209 \\
$\mu_\alpha \cos\delta$ (mas yr$^{-1}$) & $-9.681 \pm 0.072$ & $-10.061 \pm 0.083$ & $-9.669 \pm 0.016$ / $-10.078 \pm 0.019$ \\
$\mu_\delta$ (mas yr$^{-1}$) & $7.021 \pm 0.081$ & $6.256 \pm 0.095$ & $7.046 \pm 0.018$ / $6.230 \pm 0.017$ \\
$\varpi$ (mas) & $3.601 \pm 0.086$ & $3.348 \pm 0.129$ & $3.598 \pm 0.066$ / $3.330 \pm 0.080$ \\
$d_\varpi$ (pc) & $278 \pm 7$ & $299 \pm 11$ & $278 \pm 5$ / $300 \pm 7$ \\
$E(G_{BP}-G_{RP})$ (mag) & $0.104 \pm 0.027$ & $0.065 \pm 0.022$ & $0.06 \pm 0.01$ / $0.076 \pm 0.01$ \\
$A_G$ (mag) & $0.194 \pm 0.05$ & $0.121 \pm 0.04$ & $0.119 \pm 0.01$ / $0.141 \pm 0.01$ \\
$E(B-V)$ (mag) & $0.074 \pm 0.019$ & $0.046 \pm 0.016$ & $0.046 \pm 0.01$ / $0.034 \pm 0.01$ \\
{[Fe/H]} (dex) & $0.00 \pm 0.05$ & $-0.011 \pm 0.040^{*}$ & -- \\
$\mu_G$ (mag) & $7.425 \pm 0.051$ & $7.474 \pm 0.047$ & $7.221 \pm 0.002$ / $7.337 \pm 0.10$ \\
$d_{\text{iso}}$ (pc) & $279 \pm 7$ & $296 \pm 6$ & $275 \pm 1$ / $295 \pm 1$ \\
Age (Myr) & $39 \pm 5$ & $36 \pm 5$ & $28 \pm 7$ / $29 \pm 6$ \\
$(X,Y,Z)$ (pc) & $(-98,-253,-64)$ & $(-104,-271,-57)$ & $(-98,-249,-42)$ / $(-104,-270,-35)$ \\
\hline
$v_r$ (km s$^{-1}$) & $15.58 \pm 0.42$ (N=55) & $15.62 \pm 1.11$ (N=12) & $16.80 \pm 0.67^{**}$ / $16.87 \pm 0.23^{**}$ \\
$R_{gc}$ (kpc) & 8.10 & 8.11 & -- \\
$U_{\text{LSR}}$ (km s$^{-1}$) & $-12.39 \pm 0.63$ & $-9.13 \pm 0.76$ & -- \\
$V_{\text{LSR}}$ (km s$^{-1}$) & $14.05 \pm 0.34$ & $6.53 \pm 0.95$ & -- \\
$W_{\text{LSR}}$ (km s$^{-1}$) & $0.06 \pm 0.21$ & $-5.22 \pm 0.31$ & -- \\
$S_{\text{LSR}}$ (km s$^{-1}$) & $18.73 \pm 0.75$ & $12.38 \pm 1.26$ & -- \\
$Z_{\max}$ (kpc) & $0.035 \pm 0.001$ & $0.074 \pm 0.003$ & -- \\
$R_a$ (kpc) & $9.377 \pm 0.015$ & $8.534 \pm 0.073$ & -- \\
$R_p$ (kpc) & $8.192 \pm 0.002$ & $8.108 \pm 0.004$ & -- \\
$e$ & $0.067 \pm 0.001$ & $0.026 \pm 0.005$ & -- \\
Birthplace (pc) & 8544 & 8284 & -- \\
$P_{\text{orb}}$ (Myr) & $248 \pm 1$ & $233 \pm 1$ & -- \\

\hline
\end{tabular}}

\noindent\hspace{0pt}\footnotesize{${*}$: [12], ${**}$: [21]}
\end{table}

The estimated birth radii of the clusters indicate that they originated beyond the solar neighborhood, with Alessi 36 at 8544 pc and Collinder 135 at 8284 pc. Reference [20] conducted a multidimensional analysis of candidate binary systems, incorporating coordinates, trigonometric parallax, proper motion components, and color-magnitude diagrams to assess their potential physical interactions. Based on this comprehensive evaluation, the candidate systems were classified into three categories: genetic pairs (B), which share a common origin, similar age, distance, and kinematics; tidal capture or resonant trapping pairs (C), which exhibit common kinematics but may differ in age, indicating dynamic interaction without a common origin; and optical pairs (O), which appear close in the sky but are not physically associated. Although our candidate system initially appeared to resemble a genetic pair due to its similar distance, kinematics, and age, tracing the clusters' orbits back in time revealed that they originated from different locations. This finding suggests that the system does not fully match with any of the established categories, highlighting the need for a new classification to describe such cases.

\section{Conclusion}

In this study, we performed a comprehensive astrometric, photometric, and kinematic characterization of the nearby open cluster pair Alessi 36 and Collinder 135, based on the high-precision \emph{Gaia} DR3 dataset. Through a methodologically integrated approach combining with machine learning technique, color-magnitude diagram (CMD) fitting, and Galactic orbital modeling, we obtained the most probable cluster members and inferred their fundamental astrophysical and dynamical properties.

The astrometric parameters of both clusters such as proper motion components and trigonometric parallaxes were determined using the k-means clustering algorithm, focusing on stars with membership probabilities P $\ge$ 0.5. The derived mean parallaxes and proper motions were found to be in excellent agreement with previous large-scale Gaia-based cluster catalogs, such as [8]. The CMD-based isochrone fitting, performed under constraints from $Gaia$ parallaxes and spectroscopic metallicity values, yielded distance estimates of \textasciitilde279 pc and \textasciitilde296 pc for Alessi 36 and Collinder 135, respectively, along with nearly solar metallicities. Both clusters exhibit moderate but different reddening values, indicating slightly different interstellar environments despite their spatial proximity. 

The kinematic analysis utilized $Gaia$ radial velocities, confirming systemic velocities around 15.6--15.7 km s\textsuperscript{-1} for both clusters, consistent with previous spectroscopic studies. Combined with 3D astrometric parameters, we computed the full space motions and Galactic orbits of both clusters. These orbits suggest that the clusters are members of the young thin disc population, with low eccentricities and small vertical deviations (Z\textsubscript{max} < 0.08 kpc), typical for dynamically cold systems in the solar neighborhood. 

A backward integration of their Galactic orbits revealed that, although the clusters are presently co-moving and exhibit similar spatial and kinematic characteristics, they originated from distinct Galactic radii approximately 8544 pc for Alessi 36 and 8284 pc for Collinder 135. This spatial separation at birth, coupled with their slight age difference (\textasciitilde3 Myr), rules out the possibility of a common formation event. Consequently, while these clusters may appear as a binary cluster in projection and kinematics, our results indicate that they are not a genetic pair in the strictest sense. 

This finding provides an important case study in the classification of close binary open clusters (CBOCs). It highlights the limitations of conventional binary cluster classification schemes namely genetic (B), tidal resonant (C), and optical (O) by introducing an example that defies full categorization. The Alessi 36--Collinder 135 pair likely represents a dynamically coupled system without a shared origin, possibly formed via resonant trapping or convergent motion in the Galactic disc. 

Although the present analysis provides significant insights into the nature and origin of Alessi 36 and Collinder 135, several points remain open for further investigation. While this work relies on $Gaia$ DR3 radial velocities and previously published metallicities, high-resolution spectroscopic observations are needed to refine chemical abundance patterns (e.g., $\alpha$-elements, s- and r-process elements) of individual member stars. Such chemical tagging could offer crucial constraints on the clusters' birth environments and further confirm their non-genetic origin. 

As young clusters, Alessi 36 and Collinder 135 present ideal laboratories to study stellar rotation and magnetic activity. Cross-matching with large photometric surveys (e.g., TESS, ZTF) may reveal periodicities associated with stellar rotation, enabling gyrochronological age estimates as independent checks of the isochronal ages. The presence and distribution of binary systems within the clusters could impact both their internal dynamics and apparent CMD morphology. Future astrometric and spectroscopic monitoring could identify unresolved binaries and constrain the binary fraction, mass segregation, and dynamical evolution state of each cluster. 

The orbital integrations in this study were based on static Galactic potential models. Incorporating time-varying or bar-spiral-interacting potentials in N-body simulations could offer a more realistic picture of the clusters' migration history and potential capture scenarios. This may also clarify whether the clusters became a binary system after birth or were temporarily trapped in a co-moving configuration due to disc perturbations. 

In conclusion, this study illustrates the potential of $Gaia$ DR3 data not only for accurately characterizing individual open clusters but also for uncovering complex dynamical relationships among them. The findings underscore the importance of multidimensional analyses in the study of star cluster evolution and Galactic structure and pave the way for future refinements in cluster classification and formation theory.

\section*{Acknowledgements}
We sincerely thank the anonymous referees for their valuable suggestions, which greatly enhanced the quality of this paper. This study was supported by the Scientific and Technological Research Council of Turkey (T\"{U}B\.{I}TAK) 122F109. We also utilised NASA's Astrophysics Data System, along with the VizieR and Simbad databases operated by CDS in Strasbourg, France. Additionally, our study incorporated data from the European Space Agency's (ESA) \emph{Gaia} mission, processed by the \emph{Gaia} Data Processing and Analysis Consortium (DPAC), which is supported by national institutions, particularly those participating in the \emph{Gaia} Multilateral Agreement.



\nocite{*}

\printbibliography

\end{document}